# Initial stage of growth of single-walled carbon nanotubes : modelling and simulations

I. Chaudhuri, Ming Yu, C.S. Jayanthi and S. Y. Wu

*Department of Physics and Astronomy, University of Louisville, Louisville, KY 40292, USA*

**Abstract**

Through a careful modeling of interactions, collisions, and the catalytic behavior, one can obtain important information about the initial stage of growth of single-wall carbon nanotubes (SWCNTs), where a state-of-the-art semi-empirical Hamiltonian [Phys. Rev. B, **74**, 155408 (2006)] is used to model the interaction between carbon atoms. The metal catalyst forming a supersaturated metal-alloy droplet is represented by a jellium, and the effect of collisions between the carbon atoms and the catalyst is captured by charge transfers between the jellium and the carbon. Starting from carbon clusters in different initial configurations (*e.g.*, random structures, cage structures, bulk-cut spherical clusters, *etc*.), we anneal them to different temperatures. These simulations are performed with clusters placed in the jellium as well as in vacuum. We find that, in the presence of jellium, and for an optimal charge transfer of $\sim 0.1$ *e-0.2 e,* open cage structures (and some elongated cage structures) are formed, which may be viewed as precursors to the growth of SWCNTs. We will also discuss the implications of a spherical boundary on the nucleation of a SWCNT.

# I. Introduction

Single wall carbon nanotubes (SWCNTs) have attracted enormous attentions since these nanotubes were first synthesized by Ijima *et al.* in 1990 [1]. While a huge amount of fundamental research on SWCNTs, including their properties and applications, has been explored [2-17], the growth mechanism, in particular the initial stage of the nucleation and the control of the SWCNT growth, is still not fully understood.

SWCNTs can be produced by arc discharge (AD) [18], or laser ablation (LA) [19], or chemical vapor deposition (CVD) [20-22]. A transmission electron microscopy study of the CVD growth showed that SWCNTs are nucleated around small metal nanoparticles, where the diameter of the SWCNT was controlled by the size of the nanoparticle [20]. Another microscopy study showed that SWCNTs and SWCNT bundles were nucleated from larger particles [4, 6, 23, 24]. Furthermore, it is found more recently that not only the transition metal nanoparticles, but also various noble metal particles can produce SWCNTs when their sizes are of the order of a few nms [25, 26]. These observations indicate that nano-sized transition metal particles seem not to be essential for the nucleation of SWCNTs and that the nucleation simply occurs on selected nano-sized metal catalyst particles. The result thus suggests that there may be a common growth mechanism for the three growth methods based on vapour-liquid-solid (VLS) models [27]. According to the VLS model, carbon (from feedstock such as carbon-rich gases or graphite) dissolves into the liquid catalyst particle and, when the catalyst is supersaturated, carbon precipitates on the particle surface and nucleates the growth of SWCNTs. While the VLS model provides a simple explanation that captures the essence of the growth of SWCNTs, it does not provide atomistic details for the initial stage of the growth of SWCNTs.

There are also some recent experimental observations reporting the growth of nanotubes from nanodiamond [28] or semiconductor nanoparticles [29]. The CNT growth from nanodiamond particle is attributed as simply a surface diffusion and this growth mechanism is called vapor-solid surface-solid (VSSS) mechanism. Zhu et al [30] reported the growth of CNT from short nanotube-pipes and facilitated a distinction between the CNT$_{catalyst}$ and the newly formed tubes. This self-catalysis behavior of CNT may support the nucleation and radial and axial developing process.

A wide-variety of theoretical/computational studies have been attempted to unravel the atomistic details of the initial stage of the growth that include *ab-initio* molecular dynamics (MD) [4-9] , classical potential based MD [11-16], and classical thermodynamics based growth studies [31-36]. Fan *et al*. demonstrated a nucleation pathway for SWCNTs on a metal surface using a series of total energy calculations based on DFT [9]. Ding *et al*. demonstrated a nucleation process suggesting that the temperature gradient in the metal particle might be unnecessary as a driving force for nucleation of the SWCNT [11]. Raty *et al*. showed explicitly the first stages of the nucleation of a fullerene cap on a metal particle by *ab initio* MD [8]. Amara et.al [34] focused on the carbon chemical potential during SWCNT formation using tight-binding methods coupled to a grand canonical Monte Carlo (MC) simulation and showed that solubility of carbon in the outermost nickel layer is dominant in controlling the nucleation of SWNTs. Wood et al [35] proposed an analytical model for the effect of carbon flux on the growth rate of SWCNT based on a diffusion calculations. Shibuta *et* *al*. reported classical MD simulations based on empirical potentials using both Brenner [36] and Lennard-Jones potentials. They demonstrated a nanotube-cap growth process occurring on a metal cluster. This simulation emphasizes the crucial effects of temperature and size [12]. They, recently, developed a multi-scale modeling approach to study the subsequent longitudinal growth process of SWCNTs after the initial cap formation [13, 14]. These results supported various mechanisms proposed by such experimental results as 'scooter model' [7, 19], 'diffusion model' [16], and 'root growth model' [4, 6]

MD simulations are well suited for studying the nucleation mechanism of SWCNTs. However, MD schemes based on empirical potentials are not expected to provide the understanding at the quantum mechanical level for the mechanism of the nucleation of SWCNTs because they lack the framework to properly treat the bond breaking/forming processes. The quantum mechanics based *ab initio* simulations, although expected to possess the framework to predict the growth mechanism, are limited computationally. The limitation, in the case of the simulation of growth of SWCNTs, is related to two critical issues: the size of the system under consideration and the simulation time. The synthesis of SWCNTs requires metallic nanoparticles of the sizes from a few to a few tens nms. In addition, a large number of carbon atoms must be incorporated in the liquid metal droplet. It is certainly not possible to simulate such large systems with tens of thousands of atoms using first principles methods. Simulation time is another problem in quantum mechanically simulating the nucleation of SWCNTs. To overcome this issue requires an appropriate scheme to speed up the collision process to model the initial growth.

In this paper, we report our recent study on the nucleation of SWCNTs using the highly efficient simulation scheme based on a semi-empirical Hamiltonian with its framework capable of a self-consistent (SC) determination of charge redistribution and including environment dependency (ED) [37]. The framework of the SCED Hamiltonian is represented in terms of linear combination of atomic orbitals (LCAO). The SCED-LCAO Hamiltonian constructed for carbon has been shown to be transferrable for carbon-based 0-, 1-, 2-, and 3-dimensional systems [37-40], thus possessing predictive power. Our approach to the simulation of nucleation of SWCNTs is formulated to overcome the two critical issues mentioned previously so that the simulation can proceed in an efficient and appropriate manner. In Sec. II, we present a jellium-based model that is designed to capture the essential ingredients of the nucleation of a SWCNT and at the same time address the issue of the size. In Sec. III, we discuss the method of simulation, including the incorporation of the jellium potential into the SCED Hamiltonian and the scheme to speed up the collision process. Sec. IV presents the result and discussion of the simulations.

**II. The Model**

The most crucial stumbling block in simulating the nucleation of SWCNTs is the number of atoms, carbon as well as metal atoms, needed to describe the nano-sized liquid metal particles over-saturated with carbon atoms. It is not at all realistic to expect that the quantum mechanics-based simulation schemes can handle situations involving tens of thousands of atoms. However, due to the solubility of carbon atoms in these catalytic metal nano-particles (NP), carbon atoms tend to mix uniformly with metal atoms. Hence it is reasonable to assume that the liquid metal NP could be replaced by a spherical jellium medium with the carbon atoms initially distributed randomly in this medium (see Fig. 1). In this way, the size problem in the simulation is dramatically reduced.

The interactions between carbon atoms and those between carbon atoms and the jellium medium can both be described by SCED-LCAO Hamiltonians (see Sec. III). This model therefore allows an appropriate description of the collisions between carbon atoms. The catalytic effect of metal NP is intimately related to the charge transfer between metal atoms and carbon atoms. Based on the electro-negativities of relevant metal atoms and carbon atoms, it is expected that there should be, on the average, electrons transferred from metal atoms to carbon atoms. Hence, within this model, we may take into consideration of the catalytic effect and how collisions between "metal atoms" and carbon atoms affect catalysis by judiciously imposing an optimum average charge transfer while allowing the simulation to mediate individual charge transfer (see details in Sec. III). Thus the model is designed to capture the essence of the nucleation of SWCNTs via VLS, namely the effect of carbon-carbon collisions, the action of catalyst, and the catalyst carbon interactions while keeping the computation at a manageable level.

### III. Methodology

The matrix elements describing the SCED-LCAO Hamiltonian are given by [37]

$$H_{i\alpha,i\alpha}^{SCED-LCAO} = \varepsilon_{i\alpha} + (N_i - Z_i)U_i + \sum_{k \neq i}[N_k V_N(R_{ik}) - Z_k V_Z(R_{ik})] \tag{1}$$

$$H_{i\alpha,j\beta}^{SCED-LCAO} = \frac{1}{2}\{(\varepsilon_{i\alpha}^{'} + \varepsilon_{j\beta}^{'})K(R_{ij}) + [(N_i - Z_i)U_i + (N_j - Z_j)U_j]$$

$$+ [\sum_{k \neq i}(N_k V_N(R_{ik}) - Z_k V(R_{ik})) + \sum_{k \neq j}(N_k V_N(R_{jk}) - Z_k V_Z(R_{jk}))]\}S_{i\alpha,j\beta} \tag{2}$$

where $\varepsilon_{i\alpha}$ may be construed as the orbital energy of the $\alpha$-orbital of the atom at site $i$, $N_i$ the number of electrons associated with the atom at site $i$, $Z_i$ the number of valence electrons at the site $i$, $U_i$ the Hubbard-like energy representing the on-site correlation energy, $V_N(R_{ik})$ the electron-electron energy per number of electrons at site $k$ between an electron distribution at site $i$ and the electron distribution at site $k$, $V_Z$ the electron-ion interaction energy between electrons associated with the atom at site $i$ and the ion at site $k$ per number of ionic charge, $\varepsilon'_{i\alpha}$ related to $\varepsilon_{i\alpha}$, $K(R_{ij})$ a scaling function, and $S_{i\alpha,j\beta}$ the overlapping matrix. Specifically, we express $V_N(R_{ik})$ in terms of $V_Z(R_{ik})$ using a short range function $V_N(R_{ik}) = V_Z(R_{ik}) + \Delta V_N(R_{ik})$. This is because both $V_N(R_{ik})$ and $V_Z(R_{ik})$ approach $e^2/4\pi\varepsilon_0 R_{ik}$ as $R_{ik} \to \infty$.

For the simulation of the initial stage of growth of the nanotube in our model where the metal catalyst is represented by a spherical jellium, we incorporate the interaction of the jellium with the carbon atom using the jellium potential such that the Hamiltonian is now expressed as

$$H_{i\alpha,j\beta} = H_{i\alpha,j\beta}^{SCED-LCAO} + H_{i\alpha,j\beta}^{jel} \tag{3}$$

$$H_{i\alpha,j\beta}^{jel} = -\delta_{i\alpha,j\beta}(N_i - Z_i)V_i^{jel}S_{i\alpha,j\beta} \tag{4}$$

Here the jellium potential $V_{jel}^i$ is represented by

$$V_i^{jel} = \left( N_{jel}e^2 \middle/ 8\pi\varepsilon_0 R_{sph} \right)\left( 3 - R_i^2 \middle/ R_{sph}^2 \right) \tag{5}$$

where $R_{sph}$ is the radius of the spherical jellium medium and $R_i$ the distance from the center of the sphere to the $i$th carbon atom, and $N_{jel}$ the positive charge uniformly distributed in the spherical jellium medium to compensate the total net electrons transferred between all the carbon atoms and the jellium so that the charge neutrality is maintained. Denoting $n_{av}$ as the optimum average electron transferred from the jellium to the carbon atom and $N$ the total number of carbon atoms in jellium, we have $N_{jel}=n_{av}N$.

Our model is designed to capture the action of a catalyst in the initial stage of growth of a SWCNT. The action of the catalyst in our approach is modeled through the carbon-jellium interaction as already discussed. The carbon-carbon collisions, duly affected by the jellium medium, will be taken care of in the molecular-dynamics simulated annealing method based on the SCED-LCAO Hamiltonian given by Eqs. (1)-(5). The collisions between metal particles and metal/carbon atoms will be mediated by the charge transfer between the jellium medium and carbon atoms through $n_{av}$ during the annealing process. In this way, the interplay among all the interactions and the collisions will be mimicked in the modeling.

In this work, we had chosen the optimum average charge using the following approach. We placed the $C_{60}$ Fullerene in the spherical jellium. We selected $C_{60}$ because it is composed of hexagons and pentagons, geometrical elements relevant to the growth of SWCNTs. We calculated the energy of the Fullerene in the jellium medium as a function of the average charge transfer $n_{av}$ from the jellium to $C_{60}$. It can be seen from Fig. 2 that as $n_{av}$ increases from 0 (corresponding to the Fullerene in the vacuum), the total energy of the Fullerene starts to decrease while its geometrical configuration is maintained. The minimum total energy occurs at $n_{av}\sim0.1e$. Beyond $n_{av}\sim0.1e$, the energy increases, but with its geometrical shape still intact until $n_{av}\sim0.2e$. From that point on, the shape of $C_{60}$ starts to distort until it breaks apart beyond $n_{av}\sim0.25e$. We therefore chose $n_{av}=0.1e$ as the optimum average charge transfer from the jellium to carbon atoms in our investigation.

**IV. Results and discussion**

The simulation presented here is aimed to mimics the VLS model of SWNT nucleation. We have carried out a systematic study on simulating the initial stage of the growth of SWCNTs using the model presented above. Various initial configurations of carbon clusters were considered including random structures, the bucky-diamond structure, bulk-cut spherical clusters, *etc*. For the present work to test the validity of our model and approach, we have considered carbon clusters of sizes up to ~300 carbon atoms. Hence we have chosen the spherical jellium medium of diameters from 50 to 100 Å so that the carbon atoms would not escape the medium during the simulation. Fig. 3 illustrates the result corresponding to one of these initial configurations, namely a sample containing 154 carbon atoms randomly distributed in the jellium medium of a radius of 100 Å. The sample was heated to 1300 K for 5 ps to equilibrate and then slowly cooled down to 0 K. Interestingly, this process led to the formation of a tubular-like open-end structure (bottom right of Fig. 3). We found pronounced charge transfers between the medium and carbon atoms at the open-end, indicating the effect of the interaction between the metal particles and carbon atoms that plays an important role in nucleating the tubular-like structure. On the other hand, the same sample, when annealed in vacuum at the same temperature and cooled down slowly in the same manner, was observed to form a compact cluster (the bottom left of Fig. 3). This result is an indication that it is feasible to represent the role of metal catalytic particles by the jellium medium in the initial stage of growth of SWCNTs. The role played by the jellium medium is to mediate the charge transfer to/from carbon atoms, which in essence mimics the interaction between a metal catalytic particle and the carbon atoms, leading to the formation of a carbon-based tubular structure.

Fig. 4 displays the effect of jellium medium on three different types of configurations (a random configuration, a relax structure of bulk-terminated spherical diamond cluster and the bucky diamond structure). The first sample (Fig. 4a (left)), a random configuration of 68 carbon atoms, was placed in the jellium medium. It was annealed at 1500 K for 3 ps and then slowly cooled down to 0 K. A cap like hemispherical structure was observed with an open-end structure (Fig. 4a (right)). The same sample when annealed in vacuum at the same temperature and cooled down slowly in the same manner was seen to transform to a more compact structure as shown in Fig. 5a (right). The second sample is the relaxed configuration of the $C_{123}$ cluster of the bulk-(spherically) terminated diamond structure (Fig. 4b (left)). This sample was annealed at 1300 K in the jellium medium for 12 ps to stabilize and slowly cooled down. This process of annealing and cooling in the jellium medium led to the transformation of the compact structure to a tubular structure with an open end (see Fig. 4b (right)). Initially the bonds connecting the diamond core and the reconstructed surface were broken, leading to an extension of the outer shell. The atoms in the inner core of the initial compact configuration were seen to move towards the surface and an open-end cage/tubular structure was formed. The same sample when it was annealed in vacuum in the same manner and cooled down slowly was seen to form a compact structure as shown in Fig. 5b (right). The third sample, the bucky-diamond $C_{147}$ is a very stable structure where 35 atoms are in the inner core with a diamond-like structure and 112 atoms on the exterior forming a fullerene-like outer-shell (Fig. 4c (left)). This configuration was annealed at 1800 K at 5 ps and cooled down slowly in the jellium medium. A nice open-edge tubular structure was formed in the jellium medium. Initially the bonds connecting the diamond core and the fullerene-like surface were broken, resulting in an extension of the outer shell and eventually the formation of a nice open-edge tubular structure. The process was apparently mediated by the charge transfers between the jellium medium and the carbon atoms as evidenced by the large charge transfers at the open end sites. The same bucky diamond, when annealed and cooled down slowly in the same manner in vacuum, was observed to have its inner diamond core reconstructed to a shell-like configuration while the outer fullerene shell remained more or less intact, leading to an onion-like structure as shown in Fig 5c (right).

These results clearly demonstrate that the jellium atmosphere is capable of mimicking the effect of catalytic action in the formation of carbon-based tubular structures, instrumental in the initial stage of growth of SWCNTS. However, the simulation is too slow to reach the time scale required for real situations. A scheme must therefore be developed to speed up the collision process between carbon atoms. We have tested a scheme based on the introduction of an elastic spherical boundary within the jellium medium. The scheme is described in the following.

**The effect of artificial boundary**

The collision between carbon atoms will be enhanced by the introduction of an elastic spherical boundary. Fig.6 gives a schematic diagram of the situation. During the annealing, the atoms move according to the MD process. When a carbon atom hits the boundary, it will be reflected back. Hence carbon atoms will always remain inside the boundary, the rate of collision between carbon atoms will then be increased. In this way this mechanism will speed up the simulation. Let $v_x^i$, $v_y^i$, $v_z^i$ be the initial velocity components of the carbon atom along the x, y, z direction when it hits the boundary, accordingly the reflected velocities $v_x^r$, $v_y^r$, $v_z^r$ can be expressed as

$$v_x^r = \frac{1}{R^2}[(-x^2 + y^2 + z^2)v_x^i - 2xyv_y^i - 2xzv_z^i]$$

$$v_y^r = \frac{1}{R^2}[(x^2 - y^2 + z^2)v_y^i - 2yzv_z^i - 2yxv_x^i] \qquad (6)$$

$$v_z^r = \frac{1}{R^2}[(x^2 + y^2 - z^2)v_z^i - 2zxv_x^i - 2zyv_y^i]$$

where $R$ is the radius of the artificial boundary within the jellium sphere and x, y and z are the coordinate of the carbon atom. Fig. 7 shows the effect of the elastic boundary in the jellium model. For the cluster of 154 carbon atoms randomly distributed in the jellium medium (see Fig. 3 (top)), with an artificial spherical elastic boundary of radius 8 Å, the rate of collision between carbon atoms and that between carbon/metal atoms indirectly via carbon-jellium interactions are increased. The sample was annealed gradually to 1300 K for 15 ps and then slowly cooled down to 0 K. With this process, a much more pronounced tubular structure than the tubular structure obtained in the jellium medium without the boundary (see Fig. 3 (right)) was observed as shown in Fig 7 (bottom). This pronounced tubular structure may be found most likely without the boundary, but it will take a longer simulation time. This result suggests that more pronounced elongated structures could be revealed more easily with this scheme since it reduces the annealing time.

We have carried out the annealing and the cooling process for a large set of carbon clusters of different sizes and various initial configurations in vacuum, in the jellium medium, and in the jellium medium with the elastic boundary. In Fig. 8, we present the results of the simulations of a subset of four samples. These samples include three clusters $C_{87}$, $C_{99}$, $C_{123}$, with their respective initial configuration obtained from the spherical truncation of the carbon tetrahedral network, and a $C_{154}$ cluster with a random structure. The annealing for these clusters covered the temperature range from up to 1000 to 1300 K and annealing time up to 12 ps. The radius of the elastic boundary ranges from 8 to 10 Å. From Fig. 8, it can be clearly seen that the annealing of various carbon structures in the jellium medium will lead to tubular structures for these clusters while lead to more compact structures in vacuum. Furthermore the annealing will lead to more pronounced elongated structures when the annealing is carried out in jellium with an elastic boundary. In fact, this conclusion is true for all the other cases that we have considered. This is an indication that the jellium medium is capable of modeling the catalytic effect of metallic particles as long as the solubility of carbon in the metal in question would lead to a uniform mixing of carbon atoms in the metal particle. Hence the jellium medium should be an efficient and viable tool for the study of initial growth of SWCNTs

To further demonstrate the capability of the jellium medium in modeling the initial stage of growth of SWCNTs. We have studied the following scenario. We first considered a sample of 216 carbon atoms randomly distributed in a jellium medium with an elastic boundary of radius 9 Å. The annealing was carried out for 22 ps at 2000 K and then it was cooled down slowly. The final result is shown in Fig. 9a (right). It can be seen that a tubular structure had emerged, with a chain structure of some 8-10 atoms attached to the open end. This could be an indication of insufficient number of carbon atoms for continued growth. Experimentally, it is known that the continuous feedstock of carbon atoms is necessary to synthesize the SWCNTs. Hence the presence of sufficient number of carbon atoms is a necessity for the synthesis of SWCNTs. To observe whether the growth could be continued in our model, we introduced randomly 68 extra carbon atoms near the open end of the $C_{216}$ tubular structure. This new system (Fig. 8b (left)) was annealed in jellium medium in the presence of the elastic boundary at 2000 K for 24 ps to equilibrate at that temperature and then cooled down slowly. The final structure is shown in Fig. 9b (right). It can be seen that the elongated part of the structure has been lengthened, signifying the continuation of the growth, while a chain structure is again attached to the open end, indicating that further growth requires a continuation of supply of carbon atoms. Hence this demonstration provides the validation of our approach to model the initial stage of growth of SWCNTs using the jellium medium.

**V. Conclusion**

We have studied the initial stage of growth of SWCNTs by substituting the metal catalytic particle with the medium jellium and using the SCED-LCAO Hamiltonian. We have shown that the role and action of the metal catalyst in the growth process can be captured successfully by the jellium through the charge transfer between the jellium and the carbon atoms. Tubular structures with open ends have been observed to form as the result of the annealing and cooling process for various carbon clusters considered in the present work in the presence of the jellium mediated by the charge transfer between the jellium and the carbon atoms. Pronounced elongated tubular structures are formed with the introduction of the elastic boundary within the jellium medium as compared to the corresponding cases without the boundary. This is attributable to the fact that the presence of the elastic boundary increases the collision rate between carbon atoms and that between carbon atoms and metal atoms indirectly modelled through the carbon/jellium interactions. The continued growth of the tubular structure is demonstrated by the growth of the elongated portion of the tubular structure with the introduction of extra carbon atoms near the open-end of the tubular structure. It should be noted that the use of the jellium medium to substitute for the metal particles only dependent on the solubility of the carbon atoms in the catalytic metal particle to give rise to uniform mixing of the carbon atoms with metal atoms. Thus our model is consistent with recent observations that various noble metal particles of sizes of the order of a few nm can also produce SWCNTs.

We are now in the process of setting up the program based on the jellium medium to study the initial stage of growth of SWCNTs for systems of much larger numbers of carbon atoms using the order-N version of the SCED-LCAO-based simulation scheme [41,42]. We also plan to explore schemes to determine the system-specific optimum average charge transfer between carbon atoms and the system-specific jellium medium.

**Acknowledgement:** This work was supported by the US Army, Space and Missile Defense Command (Grant No.: W9113M-04-C-0024).

**Figure captions**

Figure 1   A schematic illustration of the jellium model. The grey background represents the jellium medium and the dark circles represent the carbon atoms.

Figure 2   The total energy (eV/atom) as a function of the average charge transfer $n_{av}$ from the jellium to $C_{60}$. The configurations corresponding to $n_{av}$ (0.0, 0.1, 0.2 and 0.3) are also presented.

Figure 3   A random configuration of the $C_{154}$ cluster (top)   is annealed at 1300 K and slowly cooled down to 0 K in vacuum (bottom left) and in the jellium medium (bottom right) respectively.   The open-end of the tubular like structure is marked in red.

Figure 4   Three different initial carbon configurations (left panel): (a) a random configuration of $C_{68}$, (b) a spherical bulk cut sample of $C_{123}$, and ( c) the  bucky-diamond structure of $C_{147}$ . These three samples are annealed at elevated temperature as shown in the figure and then cooled down slowly to 0 K in jellium medium.   The final configurations of these three samples are shown in the right panel. The open ends of these structures are marked in red.

Figure 5   The same samples with the same initial carbon configurations as shown in Fig. 4 (left panel) are annealed at elevated temperature and cooled down slowly to 0 K in vacuum. The final structures of these three samples are shown in the right panel. The inner cage like structure of the onion-like structure of $C_{147}$ is denoted in red.

Figure 6 A schematic diagram of the artificial boundary effect in jellium medium. The blue circle represents the artificial boundary (left).  During annealing, the carbon atoms are reflected back with reflected velocity when they hit the artificial boundary (right).  $V_i$ and $V_R$ represent the initial and reflected velocity of a carbon atom before and after colliding with the artificial boundary respectively.

Figure 7 A random configuration of $C_{154}$ (Top) is annealed at 1300 K and slowly cooled down to 0 K in jellium medium in the presence of an artificial boundary (bottom).  The open end of the tubular like structure is marked in red.

Figure 8 Three carbon clusters ($C_{87}$, $C_{99}$, and $C_{123}$) with their initial configurations cut spherically from the bulk and one random configuration of $C_{154}$ (second column) are undergoing the annealing process (third column) in three different media: vacuum (fourth column), jellium (fifth column), and jellium with an artificial boundary (sixth column), respectively. The open ends of these structures are marked in red.

Figure 9 (a) A random configuration of $C_{216}$ (left) is annealed to 2000 K and cooled down slowly to 0K in jellium medium in the presence of an artificial boundary of radius 9 Å (right). (b) Randomly 68 carbon atoms (indicated by the dashed circle) are added in the open-end  of the tubular like structure of $C_{216}$ (right) and then  the combined system is annealed to 2000 K and cooled down slowly to 0 K in jellium medium in the presence of an artificial boundary (right).

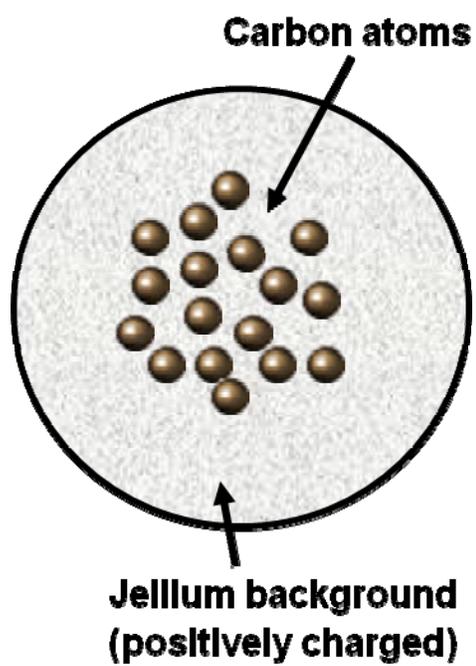

Fig. 1

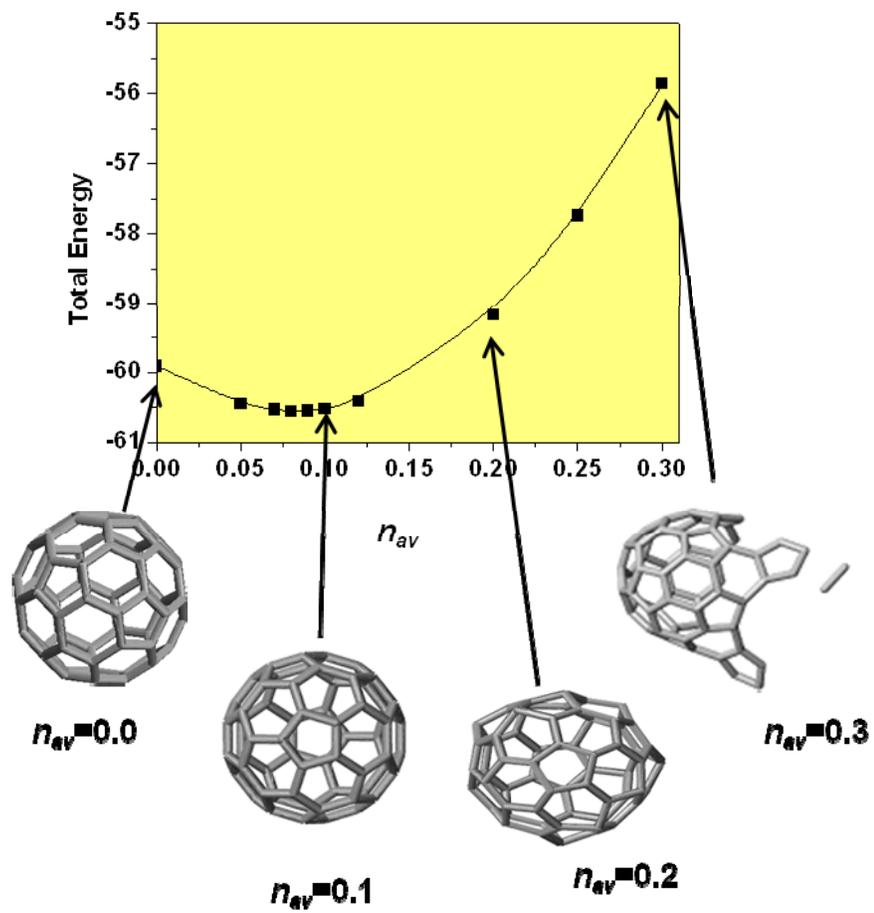

Fig. 2

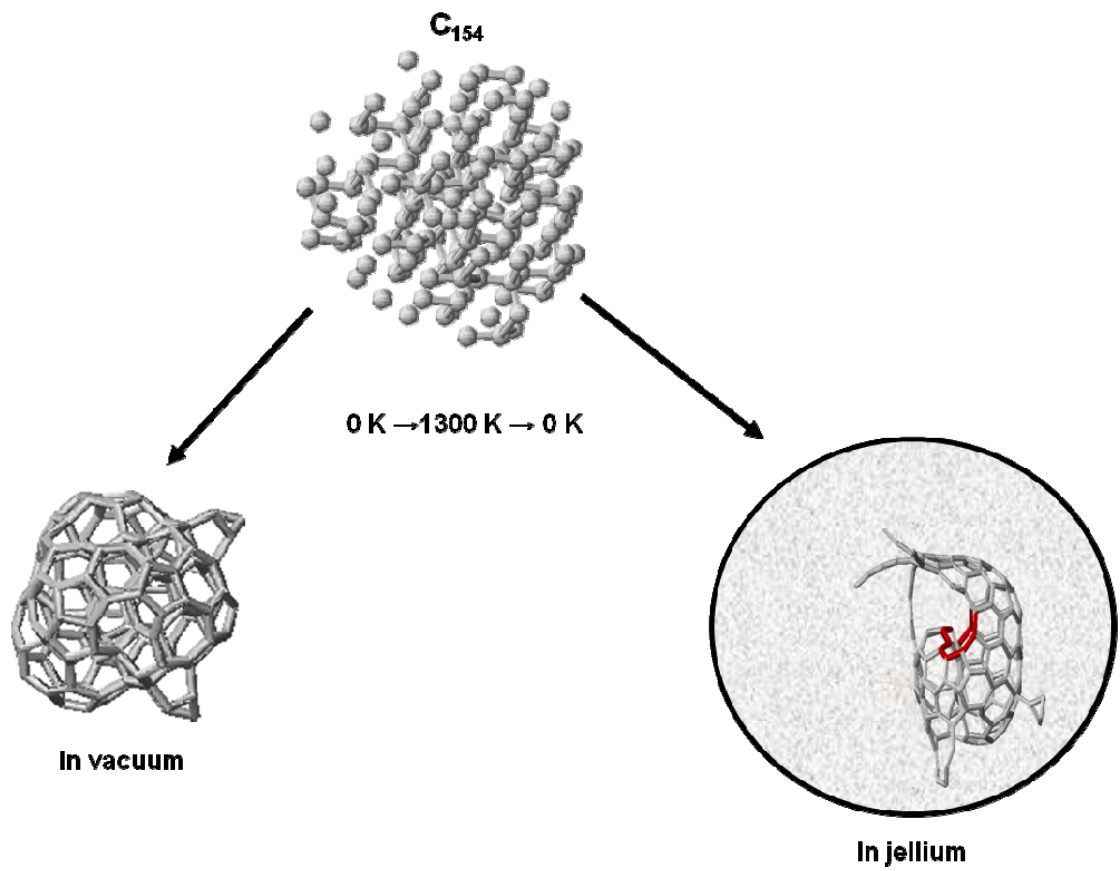

Fig. 3

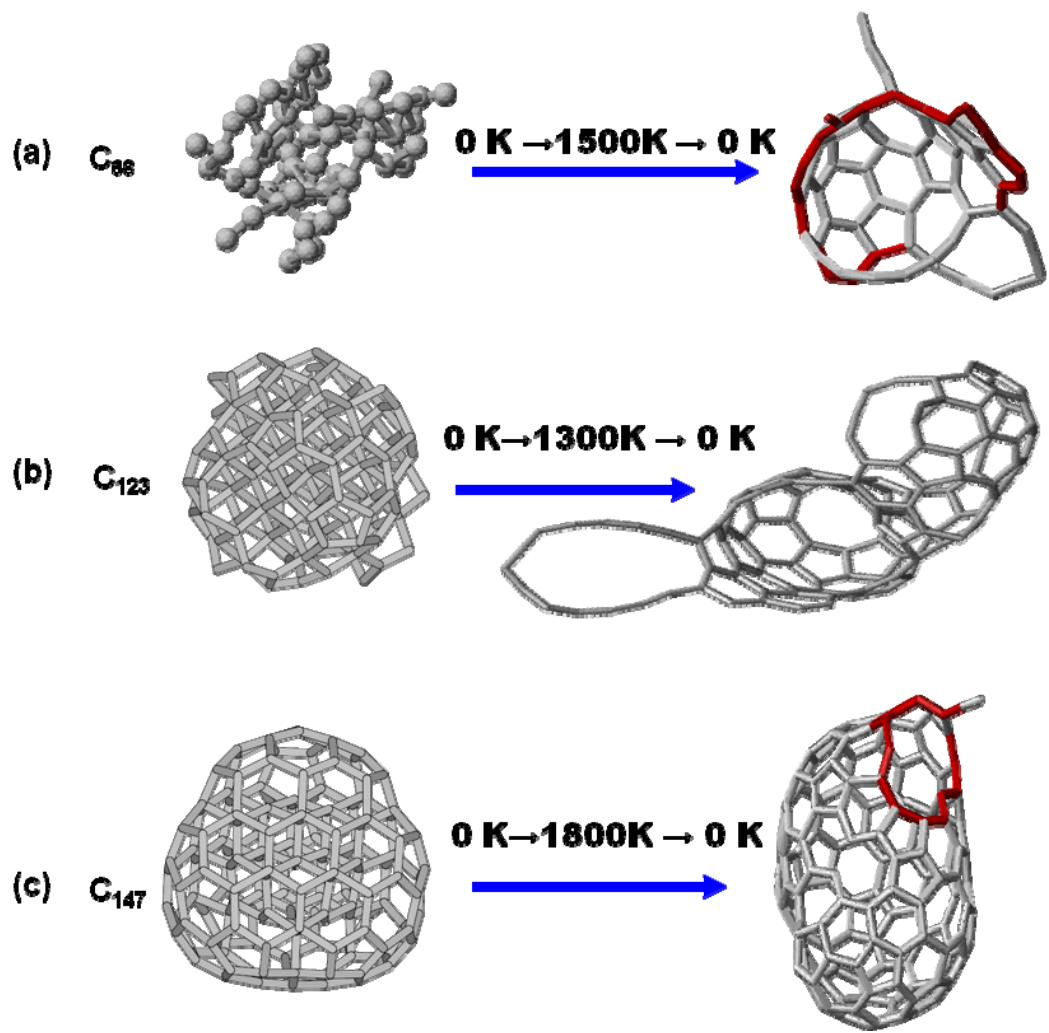

(a) C₆₈    0 K →1500K → 0 K

(b) C₁₂₃   0 K→1300K → 0 K

(c) C₁₄₇   0 K→1800K → 0 K

Fig. 4

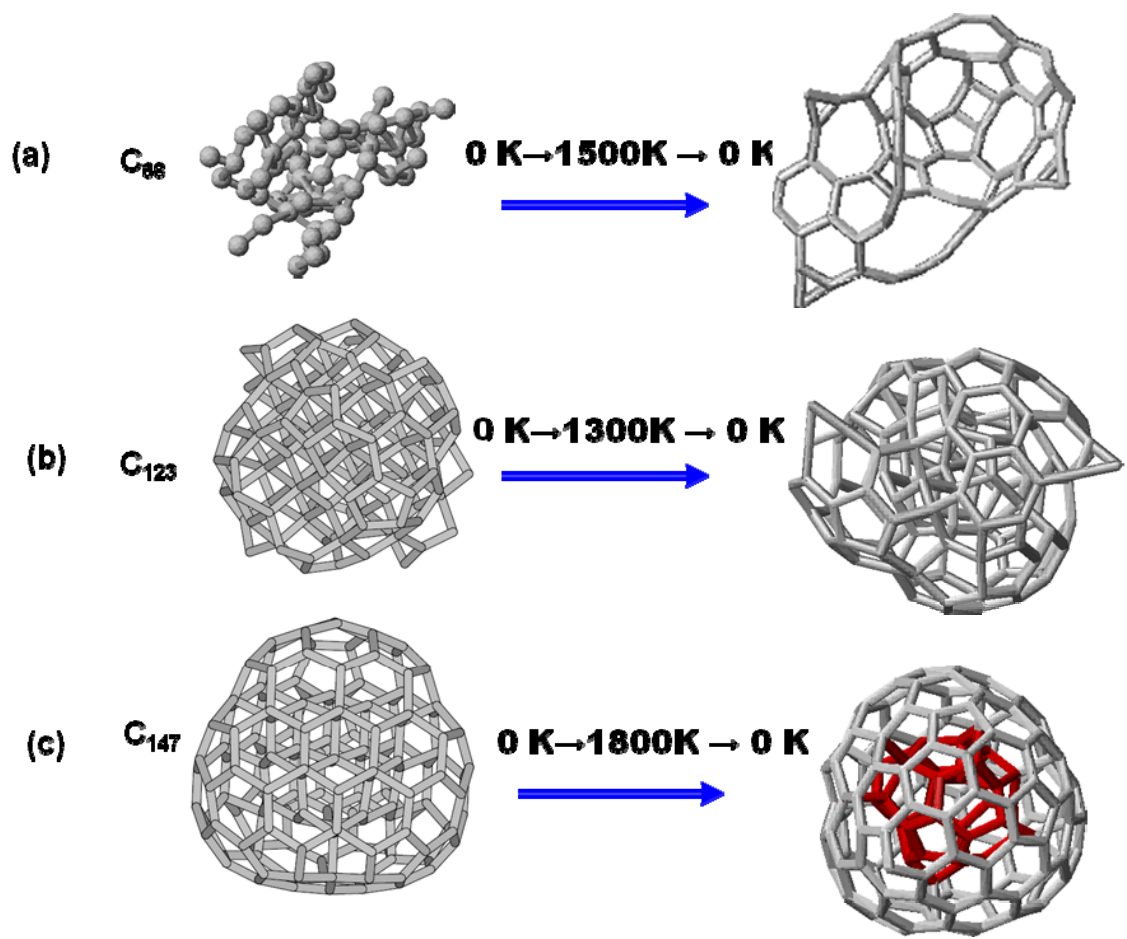

(a) C$_{88}$      0 K→1500K → 0 K

(b) C$_{123}$      0 K→1300K → 0 K

(c) C$_{147}$      0 K→1800K → 0 K

Fig. 5

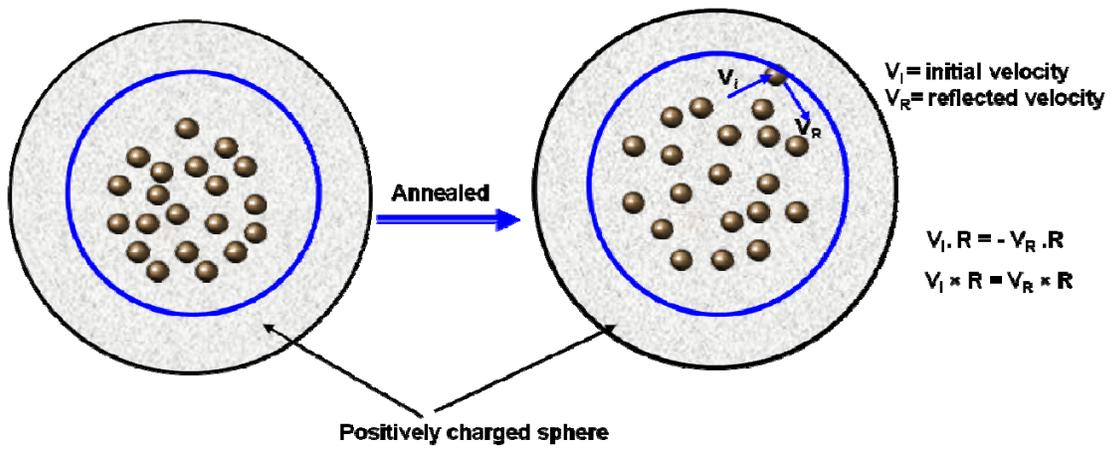

Fig. 6

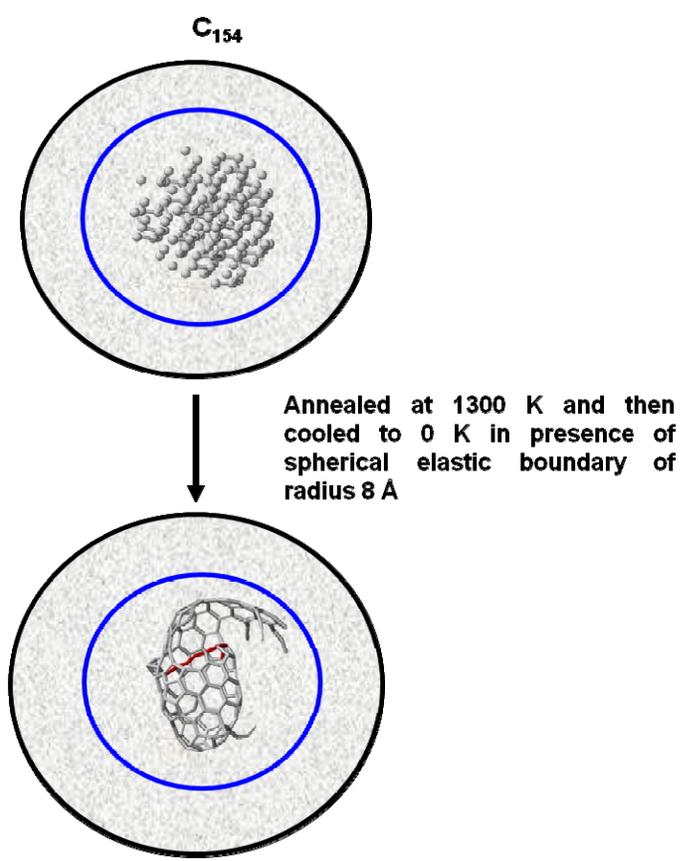

Fig. 7

| Carbon cluster | Initial configuration | Annealing process | In vacuum | In jellium medium | In jellium medium with an artificial boundary |
|---|---|---|---|---|---|
| $C_{87}$ | 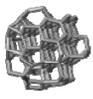 | 0K→1000K→ 0 K 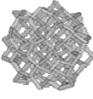 | 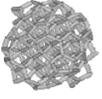 | 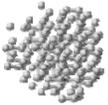 | 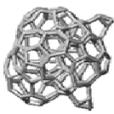 |
| $C_{99}$ | 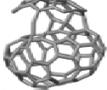 | 0 K→1300K→ 0 K 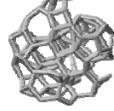 | 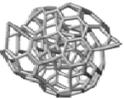 | 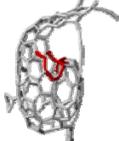 | 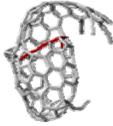 |
| $C_{123}$ | 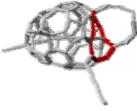 | 0 K→1300K→ 0 K 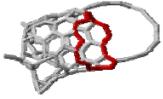 | 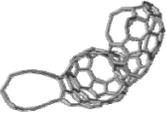 | 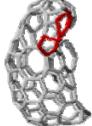 | 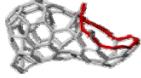 |
| $C_{154}$ | 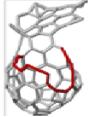 | 0 K→1300K → 0 K | | | |

Fig. 8

**C$_{216}$**

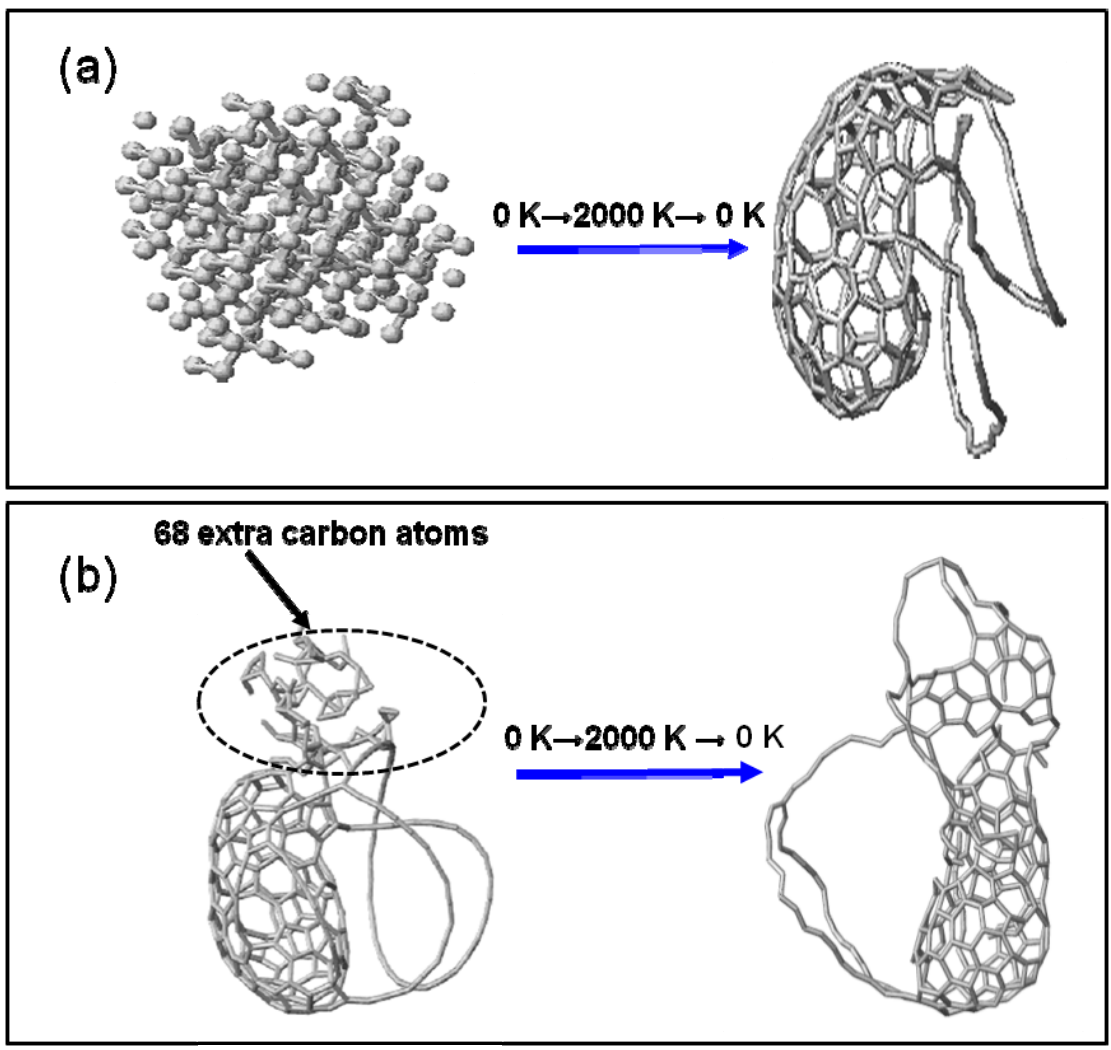

Fig. 9